\documentclass[aps,prb,twocolumn,groupedaddress,nofootinbib]{revtex4-1}

\usepackage{amsmath}
\usepackage{graphicx}
\usepackage{float}
\usepackage{tabularx}
\begin{document}
\title{Binary disorder in quantum Ising chains and induced Majorana zero modes}

\author{Jian Wang and Sudip Chakravarty}

\affiliation{Mani L Bhaumik Institute for Theoretical Physics\\Department of Physics and Astronomy, University of California Los Angeles, Los Angeles, California 90095, USA}

\date{\today}

\begin{abstract}
Dynamical structure factor \(S(k,\omega)\) is calculated for the one-dimensional (1D) transverse field Ising model, and its recent extension to include a three spin term, with quenched binary disorder. We study the low energy modes for lattices as large as 256 sites. We show that the intense zero energy modes appear whenever the binary disorder straddles two different topological winding numbers. We argue that  these are  Majorana modes, which reside  on the boundaries of the rare regions. The size distribution of Majorana pairs has a fractal behavior at the critical points. With the longer ranged interactions a spin glass  transition is  observed as well.
\end{abstract}


\maketitle

\section{Introduction}
Transverse field Ising model (TFIM) is a prototype to study quantum phase transitions~\cite{sachdev2007quantum,Dutta}. It describes a variety of quantum magnets ranging from \(\text{LiHoF}_4\)\cite{Ronnow389} to \( \text{CoNb}_2\text{O}_6 \) \cite{Coldea2010},\cite{Imai}. It is also a theoretical model to understand adiabatic quantum annealing \cite{RevModPhys.80.1061,Cirac2012,1606.07740}, where the gap closing is important.  Given Ref.~\onlinecite{1063-7869-44-10S-S29}  , TFIM can also be a play ground for topological quantum computation, where Majorana zero modes are supported at the boundaries of  1D chains. 

In the pure system, the TFIM is well understood.~\cite{Pfeuty:1970} However, disorder is an inevitable  reality. It  can come from numerous sources. Instead of universal power law  near  pure system quantum critical point,  thermodynamic properties will be highly singular in disordered systems, including log-normal distribution of gaps,~\cite{PhysRevLett.107.196804} activated  scaling, exponentially slow dynamics, and so on.~\cite{fisherPRL1992,fisherPRB1995,riegerPRB1996,youngPRB1996,guoPRL1994} 
These  can be understood as rare region effects.
There has been work on disorder effects on Majorana modes \cite{PhysRevLett.109.146403,PhysRevB.93.075129,PhysRevB.97.125408} , and the low energy distribution, but little is known about the spatial distribution of the Majorana modes, and their relation with Griffiths-like rare regions.
In this paper, we will clarify the relation of  rare regions,  exponentially slow dynamical excitations, and the exponential degeneracy of  Majorana zero modes.

The outline of this paper is as follows: in section \ref{section:methods}, the problem is stated quite generally, including  the three spin interactions studied recently, and the method of calculating spin-spin correlation function, and the dynamical structure factor. 

In section \ref{section:2-spin_binary}, the condition for the emergent low energy states  is given for the TFIM for the binary transverse field disorder that gives rise to the interesting effects discussed in the present paper. We do not consider other forms of disorder such as the box or the Gaussian distribution.The  binary distribution consists of  a large field \(h_L\) and a small field \(h_S\), with probability \(P_L+P_S=1\) such that 
\begin{eqnarray}
h_i=\begin{cases}
h_L   \qquad \text{Probablity}  =P_L := P\\
h_S  \qquad \text{Probablity} =P_S 
 
\end{cases}
\end{eqnarray}

In section \ref{section:2-Griffith}, we will argue that these low energy modes are Majorana zero modes (MZM) at the boundaries of Griffiths-like rare regions by correlating the spin-spin correlation function and the lowest energy Majorana eigenvectors. The size distribution of MZM is also calculated at the critical point;  the most probable size is $1/5$ of the system length; we also notice a self similar distribution.

In the Section~\ref{lrange}, disordered longer ranged interaction is explored.~\cite{PhysRevB.85.035110} (see also Ref.~\onlinecite{DeGottardi}) The rare region induced Majorana zero mode picture is similar to the TFIM.This is a model  that exhibits a very rich phase diagram. Given  frustration from the longer ranged interaction, we also note a spin-glass phase transitions in this case.

\section{the Hamiltonian}
\label{section:methods}
The transverse field Ising chain with longer range interaction with disorder~\cite{PhysRevB.85.035110} is
\begin{eqnarray}
	H=-\sum_{i=1}^{L} h_i \sigma^z_i -\sum_{i=1}^{L-1} \lambda_{1i} \sigma^x_i \sigma^x_{i+1}  -\sum_{i=1}^{L-2} \lambda_{2i} \sigma^x_i\sigma^z_{i+1} \sigma^x_{i+2} \qquad
	\label{eq:3spinHamiltonian}
\end{eqnarray}
where \(i\) is the site index and \(L\) is the size of the 1D system with open boundary condition. Here \(h_i\) is quenched transverse field, \(\lambda_{1i}\) are two-spin couplings and \(\lambda_{2i}\) are three-spin couplings,  they can be of constant value, or assume random variables satisfying  certain distributions.

The spin-spin correlation function, is calculated using the ground state average \( \langle \cdots \rangle \) and  the disorder ensemble average is denoted by an overbar
\begin{eqnarray}
 C(r,t):= \overline{\langle \sigma^x_i(t)\sigma^x_j(0) \rangle}
 \label{eq:correlator}
\end{eqnarray}  Since our disorder averaged system is translationally invariant, we use $r$ as the distance between two sites.
The dynamical structure factor \(S(k,\omega)\) is the time and spatial Fourier transformation of the spin-spin correlation function:
\begin{eqnarray}
	S(k,\omega)=\int dt \int dr \ e^{i\omega t} e^{-ikr} C(r,t)
	\label{eq:Fourier}
\end{eqnarray}

\subsection{Jordan Wigner transformation}

From  Jordan-Wigner transformation~\cite{Pfeuty:1970} the system can be expressed in terms  of single-particle  fermion operators to solve the eigenvalues and eigenvectors for a given lattice with a given disorder configuration

\begin{widetext}
\begin{eqnarray}
H=\sum_{i=1}^{L}h_i (c^\dagger_i-c_i)(c^\dagger_i+c_i)-\sum_{i=1}^{L-1}\lambda_{1i} (c^\dagger_i-c_i)(c^\dagger_{i+1}+c_{i+1}) -\sum_{i=1}^{L-2}\lambda_{2i} (c^\dagger_i-c_i)(c^\dagger_{i+2}+c_{i+2})  \qquad
\label{eq:fermionHamiltonian}  
\end{eqnarray}
\end{widetext}
The spin-spin correlation function in terms of fermion operators is given by 
\begin{equation}
 \langle\sigma^x_i(t) \sigma_j^x(0)\rangle=\langle (c^\dagger_1(t) +c_1(t)) \cdots  (c^\dagger_j(0) +c_j(0) \rangle.
 \end{equation}  
Using Wick's theorem (see, for example, Ref.~\onlinecite{sachdev2007quantum}), the right hand side can be expressed as  a Pfaffian of a \(2i+2j-2\) dimensional matrix. Each element in the matrix is  a free two-fermion correlator.

\subsection{Pure system}

\begin{figure}
	\includegraphics[width=0.8\linewidth]{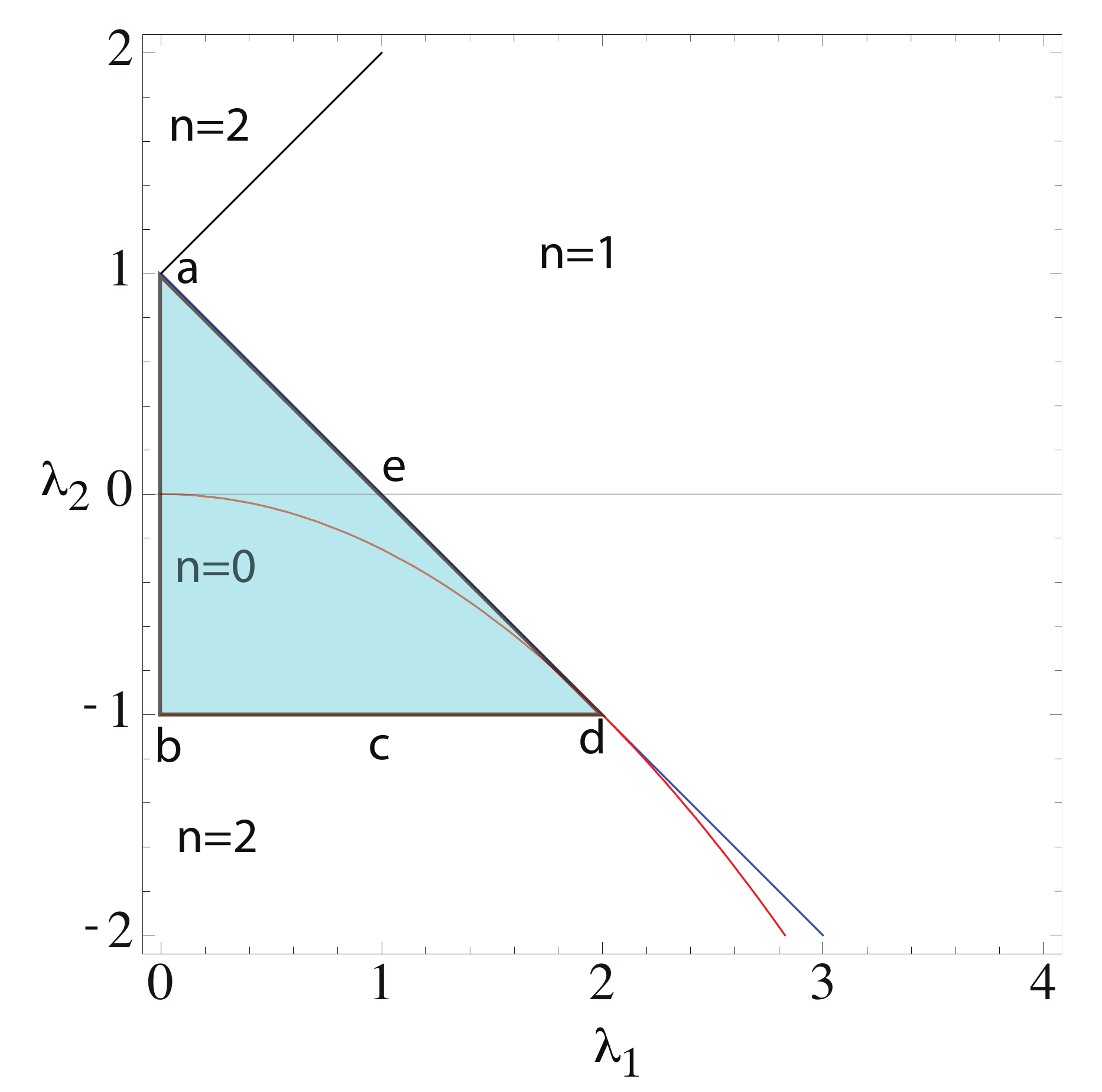}%
	\caption{\label{fig:phasediagram}  Pure system phase diagram of 3-spin model $H$.  The transverse field is taken to be unity.  The labels \(n=0,1,2\) are the topological numbers, denoting pairs of Majorana modes at open boundaries.~\cite{PhysRevB.85.035110} }
\end{figure}

Figure \ref{fig:phasediagram} is the  phase diagram of the pure 3-spin model.~ \cite{PhysRevB.85.035110} The transverse field \(h\) is set to unity. The horizontal line at \( \lambda_2=0\) corresponds to the transverse field, with critical point at \(e\). In this phase diagram, the \(n=1\) regions correspond to the magnetically  ordered regions.

\begin{figure}{htb}
	\includegraphics[width=\linewidth]{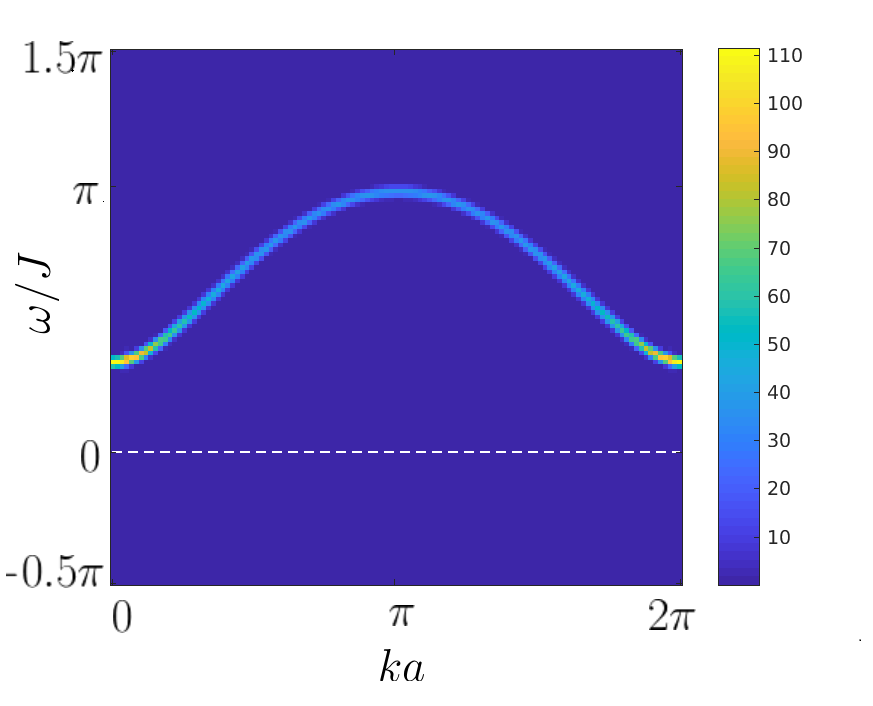}%
	\caption{ the dynamical structure factor
		120 sites, paramagnetic phase: \( h=1 \quad \lambda_1=0.5 \quad \lambda_2=0 \) }
	\label{fig:Swk101}
\end{figure}

\begin{figure}{htb}
	\includegraphics[width=\linewidth]{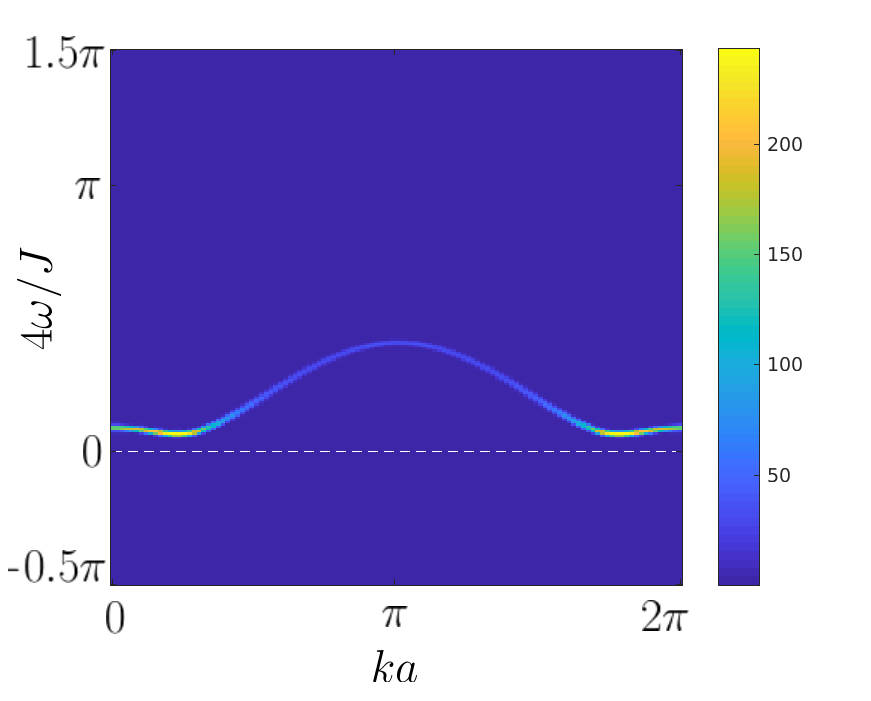}%
	\caption{ Dynamical structure factor of pure system \(h = \lambda_1 = 1   \quad \lambda_2 = -0.5\), 120 sites
		the dispersion curve has a dip at non-zero \(k\) value, that gap can also be closed at non-zero
		\(k\) by tuning  parameters}
	\label{fig:Swk103}
\end{figure}

\begin{figure}{htb}
	\includegraphics[width=\linewidth]{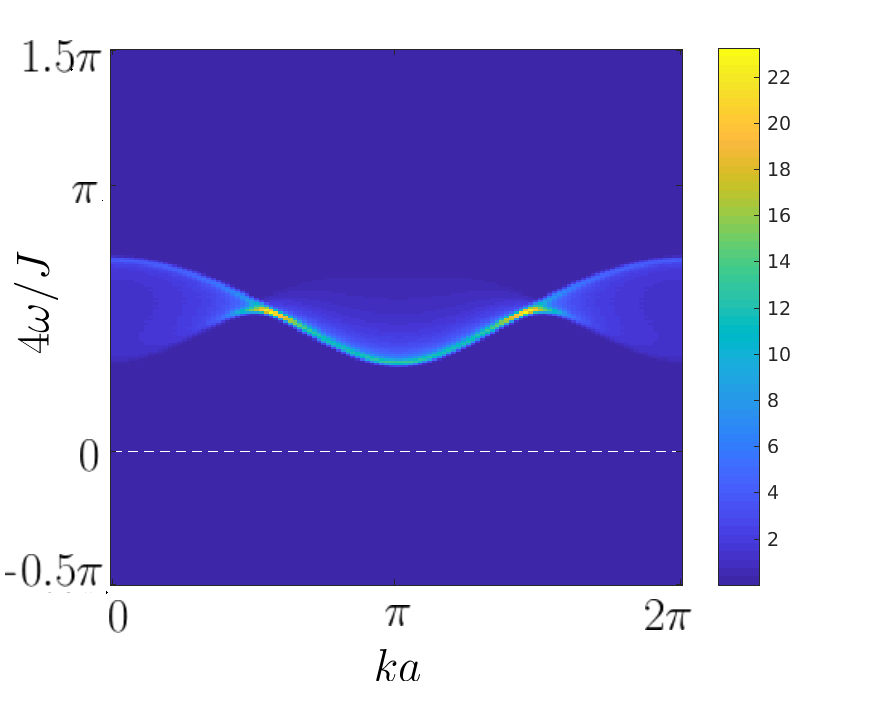}%
	\caption{ Dynamical structure factor of pure system   \(h = \lambda_1=\lambda_2 = 1 \), 120 sites
		this is in the two particles continuum region, single spinon excitation is forbidden in this
		n = 1 phase	\label{fig:Swk102}  }
\end{figure}

To explore the dynamics in the phase diagram, we plot a few examples of \( S(k,\omega) \):  Fig.~\ref{fig:Swk101}   \( (h=1,\lambda_1=0.5,\lambda_2=0) \) ; Fig.~\ref{fig:Swk103} \( (h=1,\lambda_1=1,\lambda_2=-0.5) \); Fig.~\ref{fig:Swk102} \( (h=1,\lambda_1=1,\lambda_2=1) \)

Note that Fig. \ref{fig:Swk102}  is in the magnetically  ordered region \(n=1\); \(m^2\) has been subtracted from the spin-spin correlation function, and the excitation is two-particle continuum.

\section{Emergent low energy modes in disorder chain	}
\label{section:2-spin_binary}

\begin{figure}[htb]
  \includegraphics[width=\linewidth]{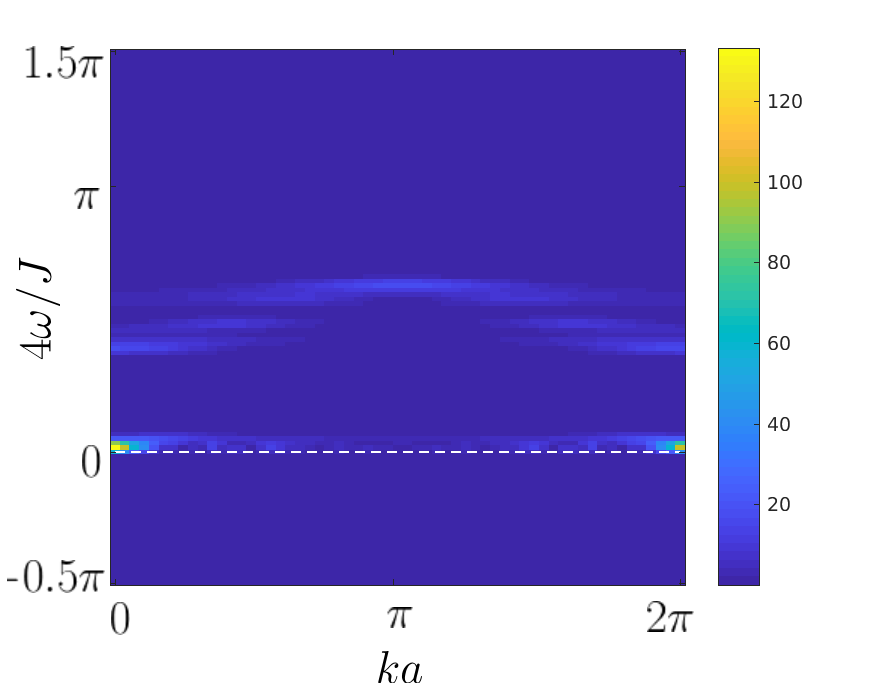}%
  \caption{\label{fig:Swk} \(S(k,\omega)\) of a 2-spin model: the quenched disorder transverse field \(h_i\) satisfies binary distribution with \(P(h_i=3.0)=0.6   \)     and \( P(h_i=0.2)=0.4 \) (120 sites).}
  \end{figure}

In this section, let's only consider the 2-spin Hamiltonian  \(H_{2}=-\sum_{i=1}^L h_i \sigma^z_i -\sum_{i=1}^{L-1} \sigma^x_i \sigma^x_{i+1} \)   with \( \lambda_1=1 \quad {\rm and}\;  \lambda_2=0 \), the random transverse field has the binary distribution:  the larger field \(h_L\) and the smaller field \(h_S\), with probability \(P_L+P_S=1\)
As $P$ is changed  from $0$ to $1$, we will show that, for \( 0<h_S<1<h_L \)  there is a phase transition as we change $P$, and there will be low energy emergent  modes.
Consider, for eample, $h_L=3.0, P_L=0.6, \text{and}\;  h_S=0.2, P_{S} = 0.4$
For these parameters we get the spectra shown in Fig. \ref{fig:Swk}. The spectral density has a very strong peak near the zero energy \(\omega=0 \), and near \(k=0\). At higher energies, the spectra breaks up into horizontal stripes. The central question of this paper is to understand what is the origin of the low energy signal.

\begin{figure}[htb]
	\includegraphics[width=1.1\linewidth]{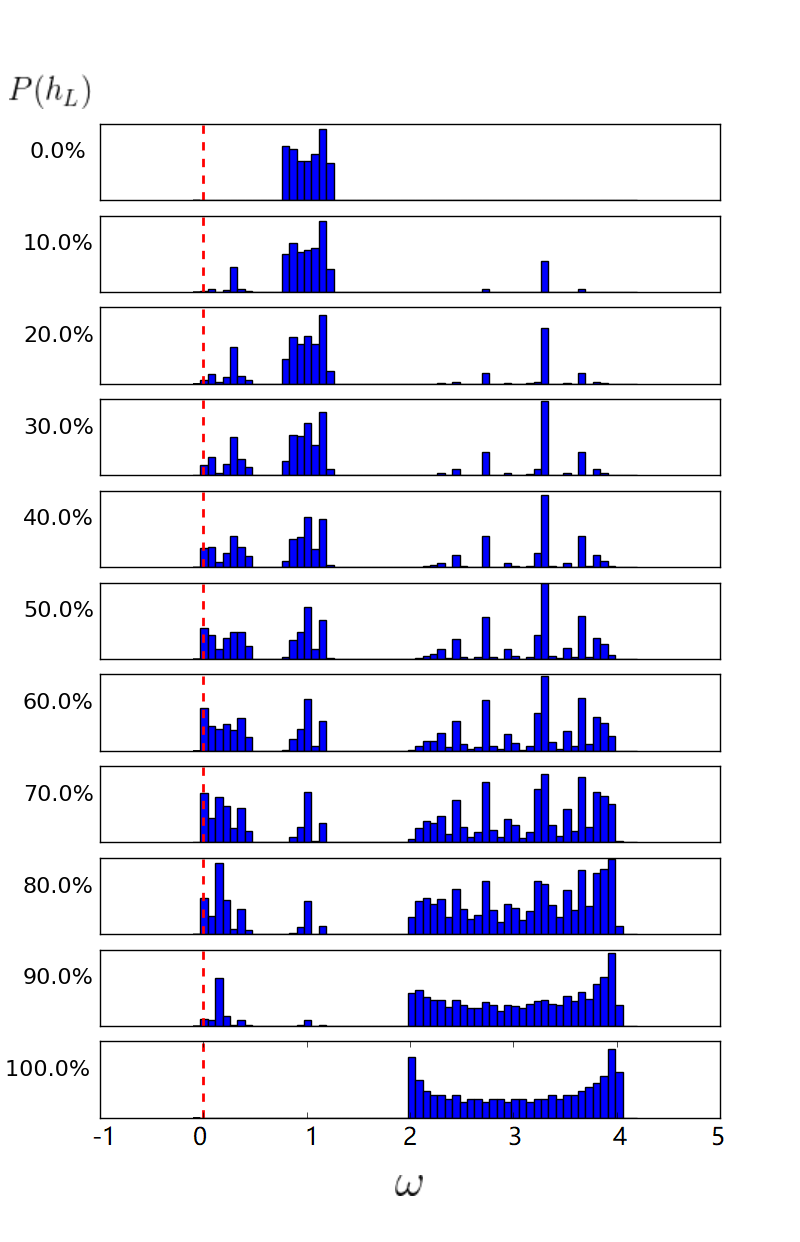}%
	\caption{\label{fig:dos}  11 density of state plots for system with binary transverse field disorder. All of the graphs have the same large field \(h_L=3.0\) and small field \(h_S=0.2\), the difference is the binary disorder. The two extreme density of states plots, on the top \(P=0\%\) and in the bottom \(P=100\%\), correspond to the pure system in ferromagnetic and  paramagnetic phases.  From top to bottom, the probability of large field is increasing, the probability of small field is decreasing  }
\end{figure}

 To answer the question, let's fix \(h_S\) and \(h_L\), and take \(P\) as a tuning parameter. Then it can be converted into another question: how do the zero energy modes emerge as a function of \(P\)? 
We plotted the density of states  in Fig.~\ref{fig:dos}. There are 11 graphs for different \(P\) values. The  \(P_L=0.6\) corresponds to  Fig.~(\ref{fig:Swk}). Notice that the density of state \(\rho(\omega) \)  and the integrated \( \int S(k,\omega) dk \) are related.

\begin{figure}[htb]
	\includegraphics[width=1.0\linewidth]{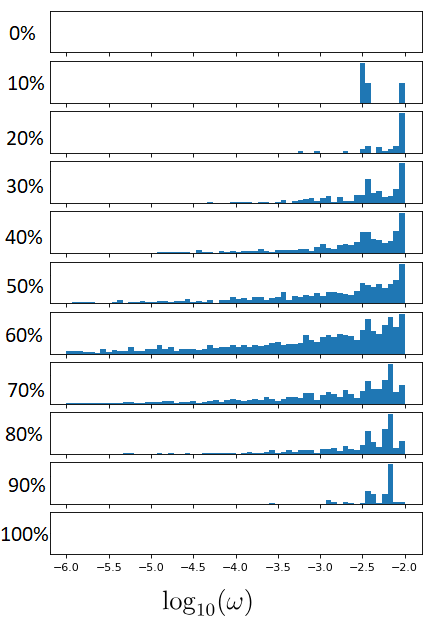}%
	\caption{\label{fig:logDOS}  The density of state plot near zero energy of Fig.~\ref{fig:dos}.  The horizontal axis is in \( \log_{10}\) scale. This detailed study shows that, although the gap looks closed for all disorder  in Fig.~\ref{fig:dos}, there is a optimal percentage, where the closing is the best  }
\end{figure}

In Fig.~\ref{fig:dos},  the two extreme cases \(P=0 \) and \(P=1\) are gapped, with no zero energy modes. For intermediate values, we can see the zero energy modes.
In Fig.~\ref{fig:logDOS}, we plot the density of states near the zero energy, on a log-scale. It capture the details about how the gap is closed. 

From Fig.~\ref{fig:dos}, we can also find that the excitations are grouped into three regions. 
\begin{enumerate}
\item \( h_L-1 <\omega < h_L+1  \) corresponds to the excitations in the paramagnetic region.
\item \( 1-h_S <\omega < 1-h_S  \) corresponds to the ferromagnetic phase. The two-particle continuum excitations is not obvious in \(S(k,\omega)\) graph.
\item \( \omega < \omega_0  \)  corresponds to the emergent low energy modes.
\end{enumerate}
The energy is always bounded by these groups, no matter what the disorder is.

The emergent low energy modes do not always exist. By tunung  \(h_L, h_S\), we find:
\begin{itemize}
	\item it exists when the large  and the small fields straddle the critical point \( 0< h_S < 1 <h_L \).  For the cases of \(   1<h_S <h_L \)  or  \(  0< h_S <h_L<1 \) , no low energy mode  emerges, no matter what \(P\) is.
	\item in the proper case \( 0< h_S < 1 <h_L \),  there is a value of \(P\) which generates maximum numbers of low energy modes, and the gap is minimized. We will show that such a point is  given  by \( \overline{\ln h_i} =\overline{\ln J_i} \) \cite{PhysRevB.53.8486}
\end{itemize}

\subsection{the critical point}
In the two extreme cases in Fig.~\ref{fig:dos}, i.e. with no disorder, \(P=0.\) represents the ferromagnetic phase, and \( P=1\) represents the paramagnetic phase.
At an intermediate value of \(P_L\)\, the system must have a quantum phase transition.

The critical value of \(P_C\) is given by:
\begin{eqnarray}
     \overline{\ln h_i}=\overline{\ln J_i } \nonumber \\
	h_L^{P_C} h_S^{1-P_C}=1 \nonumber \\ 
	P_C=\frac{\ln h_S}{\ln h_S - \ln h_L} \label{eq:Pc}
\end{eqnarray}

\begin{figure}[htb]
	\includegraphics[width=1\linewidth]{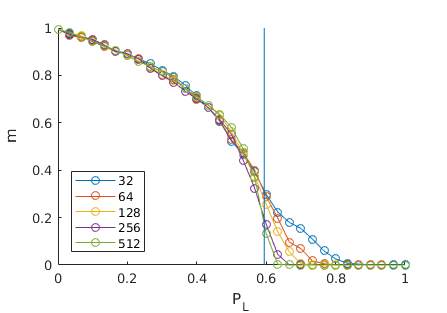}%
	\caption{\label{fig:magne} The magnetization as a function of binary distribution parameter \(P\) , the random transverse field \(h_i = P\delta(h-h_L) +(1-P)\delta(h-h_S)\) ,  \(h_L=3.0 \ h_S=0.2 \)  we can see the critical behavior predicted by  \(P_C=\frac{\ln h_S}{\ln h_S - \ln h_L}=0.59 \)}
\end{figure}

In Fig.~\ref{fig:magne}  the magnetization  is plotted, for \( h_S=0.2 \quad \textrm{and} \; h_L=3.0 \). The vertical line is the critical value \(P_C=\frac{\ln 0.2}{\ln 0.2 -\ln 3.0}\approx 0.5943 \). The magnetization is calculated by \( m= \sqrt{ \overline{\langle \sigma^x_{+\infty}(0)\sigma^x_0(0) \rangle} } \), with 129 random configurations for the ensemble average.
Due to the randomness, the magnetization takes large computational resources in the  averaging procedure. 

\begin{figure}[htb]
	\includegraphics[width=1\linewidth]{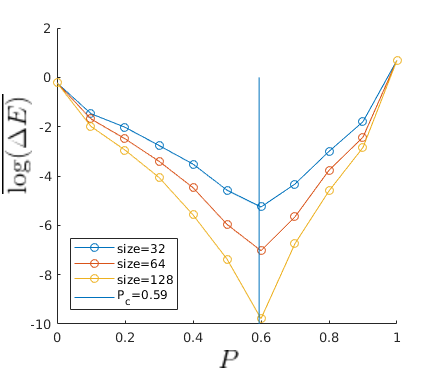}%
	\caption{\label{fig:Egap} The log of energy gap as a function of binary distribution parameter \(P\) , the random transverse field \(h_i= P\delta(h-h_L) +(1-P)\delta(h-h_S)\) . At  \(h_L=3.0 \ h_S=0.2 \)  we can see the critical behavior predicted by  \(P_C=\frac{\ln h_S}{\ln h_S - \ln h_L}=0.59 \). Notice that the ensemble average is the typical average, it is the \(\text{mean}\{\epsilon_{gap_i}\}\), not the \(\text{min}\{\epsilon_{gap_i}\}\)}
\end{figure}

The Fig.~\ref{fig:Egap} is the energy gap plotted against \(P\). In our calculation, we choose periodic boundary condition for the fermions.  We also choose logarithmic scale for the energy. Without the log-scale, they all look close to zero; see   Fig.~\ref{fig:dos}.

\begin{figure}[htb]
	\includegraphics[width=1\linewidth]{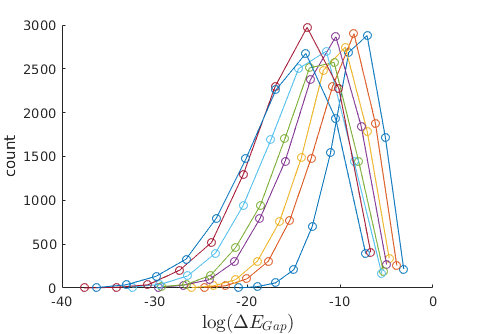}%
	\caption{\label{fig:Esize} The gap distributions for different system sizes. horizontal axis is the log of the energy gap \(\log(\Delta E)\); the vertical axis is the distribution count. The system sizes from right to left are \(L=32,48,64,80,96,112, 128,114\); the random average consisted of 10000 random samples}
\end{figure}

\begin{figure}[htb]
	\includegraphics[width=1\linewidth]{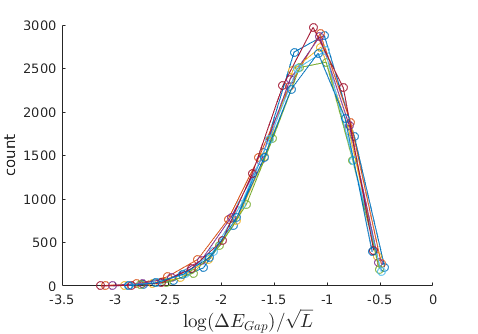}%
	\caption{\label{fig:Esize1} The collapse of the data in Fig.~\ref{fig:Esize}. The horizontal axis is rescaled by the square root of the system size. This collapse only happens at the critical point \(P_C=0.59\)}
\end{figure}

The Figs.~\ref{fig:dos}, ~\ref{fig:magne}, and ~ \ref{fig:Egap} have already shown that, the critical point exists, and it is predicted by Eq.~ (\ref{eq:Pc}). 
The Figs. \ref{fig:Esize} and ~ \ref{fig:Esize1} demonstrate the activated scaling at the quantum critical point; the energy gap is proportional to  \( e^{-\alpha \sqrt{L}} \)

\section{Griffiths-like phase and the Majorana zero modes}
\label{section:2-Griffith}

In the last section we have shown that there is a phase transition as a function of $P$, and the low energy modes emerge close to the critical point.
In this section, we will explore the nature of the low energy modes.

\subsection{Disorder induced rare regions}

We know that in a pure system, Majorana modes exist at the  boundaries  of a topologically non-trivial phase.
In a disordered system, which is not uniform, it is possible that a spatial region is in the non-trivial phase, while the surrounding region is still in the trivial phase. Thus the low energy Majorana zero modes are created by rare  regions of magnetization,  the ``Griffiths phase''.

To understand, let's plot the equal time spin-spin correlation function  for a specific random configuration:
\begin{eqnarray}
<\sigma^x_n \sigma^x_m>
\end{eqnarray}
\(n\) and \(m\) run from \(1\) to \(L\); so this plot contains the correlation of each pair at equal times. Here are some important properties:
(1) the diagonal term is always unity, \(<\sigma^x_n \sigma^x_n>=1\); (2) it is symmetric under \( m \leftrightarrow n \); (3) it is real because, \(<\sigma^x_n \sigma^x_n>^*=<\sigma^x_n \sigma^x_n> \)

\begin{figure}[htb]
	\includegraphics[width=1\linewidth]{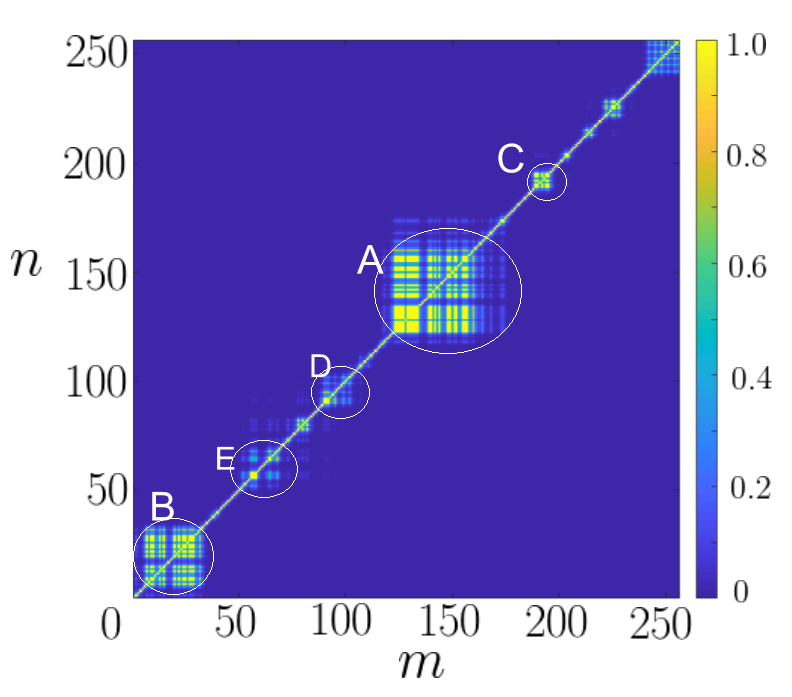}%
	\caption{\label{fig:MZMrareRegions} equal time spin-spin correlation \( <\sigma^x_n \sigma^x_m> \) ,  the horizontal axis is \(m\) the vertical axis is \(n\), the color is the strength of \(<\sigma^x_n \sigma^x_m> \) }
\end{figure}

From Fig.~\ref{fig:MZMrareRegions}, one can see the rare regions clearly by watching which site is correlated with which site.
It  is a spin-spin correlator plot:  \(h_L=3.0\) with \(60 \%\) probability and \(h_S=0.2\) with 40\% probability. We can see the cluster of rare regions A,B,C,D,E.
The largest region A spans about 30 sites from 125 to 160, it is where the small field \(h_S=0.2\) are gathered. Since the field is weak there, the spins tend to be coupled by interaction, and correlated to form magnetic order. Although, at certain sites, the cluster may contain large field, the cluster is not broken by it. At a coarse grained level, it is single giant spin.

\begin{figure}[H]
	\includegraphics[width=1\linewidth]{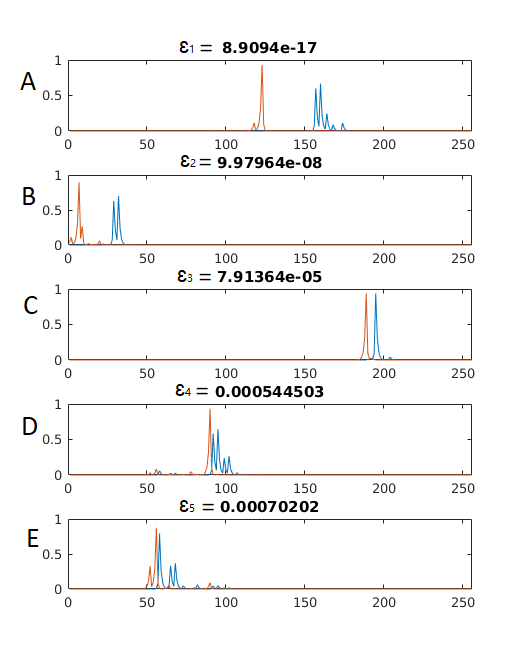}%
	\caption{\label{fig:MZMs} Five lowest eigenvectors, titles are the energy  \(\epsilon_n \), index \(n\) is in ascending order of eigen-energy. The horizontal axis is the lattice site, the vertical axis is the value of \( \psi_{ni} \) and \(\phi_{ni}\). The orange and blue curves correspond to decoupled Majorana pairs, the  real part \( \tilde{a}_n=\sum_i\psi_{ni} a_i \) and  the imaginary part \( \tilde{b}_n=\sum_i\phi_{ni} b_i \).  A,B,C,D,E  correspond to the rare regions in Fig.~\ref{fig:MZMrareRegions}  }
\end{figure}

The quadratic fermion Hamiltonian in Eq.~\ref{eq:fermionHamiltonian}, can also be rewritten in the of Majorana representation:
$a_i =c^\dagger_{i}+ c_i$ and   $b_i = i(c^\dagger_{i}- c_i )$.
The Hamiltonian is then
\begin{eqnarray}
	H=i \sum_{i=1}^{L}h_i  b_i a_i+i\sum_{i=1}^{L-1}\lambda_{1i} b_i a_{i+1} +i\sum_{i=1}^{L-2}\lambda_{2i} b_i a_{i+2} \qquad
	\label{eq:majoranaHamiltonian}
\end{eqnarray}
The Equation \ref{eq:majoranaHamiltonian} can be solved with singular value decomposition, into  decoupled Majorana pairs:
\begin{eqnarray}
H=i \sum_{n=1}^L \epsilon_n \tilde{a}_n \tilde{b}_n
\end{eqnarray}
The Fig.~\ref{fig:MZMs} shows the five lowest eigenvectors of  \( \tilde{a}_n=\sum_i\psi_{ni} a_i \) and \( \tilde{b}_n=\sum_i\phi_{ni} b_i \), the vertical axis labels the eigenenergy \( \epsilon_n=\Lambda_{nn} \).  

These Majorana pairs are the eigenstates representing  the many-body  excitations. From top to bottom, those Majorana pairs in Fig.~\ref{fig:MZMs} correspond to the rare regions A,B,C,D,E in Fig.~\ref{fig:MZMrareRegions}

\begin{itemize}
\item  Majorana pairs reside at the boundary of magnetic rare regions.
\item if the magnetic rare region's boundary is not sharp, the Majorana mode will span a large distance
\item when the Majorana pairs  get closer, their energy increase
\end{itemize}

\subsection{The separation between the Majorana zero mode pairs}
The separation of a Majorana pair is defined by:
\begin{eqnarray}
	s_n=|\frac{\sum_i i |\psi_{in}|^2}{\sum_i |\psi_{in}|^2} -\frac{\sum_i i |\phi_{in}|^2}{\sum_i |\phi_{in}|^2} |
	\label{eq:MZMsize}
\end{eqnarray}
the \(n\) labels the different eigenmodes; \(i\) is the lattice site. The above definition works for any Majorana  eigenvectors (not necessarily the zero mode) . But we are  interested in the behavior of the low energy modes, because for high energy modes, \(\psi_{in}\) and \(\phi_{in}\) will significantly overlap, and \(s_n\) will be trivially small.

We define \(n=1\) to be the lowest energy mode (eigenvalues are in ascending order). Then \(s_n\) will be the largest  separation distance.  We plot the distribution of relative sizes, \(s_n/L\), for a random ensemble, at the critical point.

 \begin{figure}[H]
	\includegraphics[width=\linewidth]{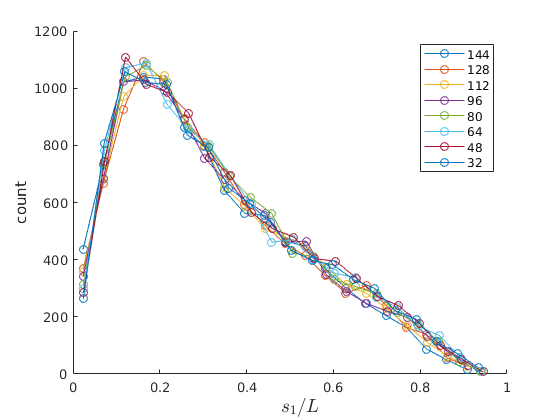}%
	\caption{\label{fig:size1}  The distribution of the sizes of Majorana pairs at the critical point \(P=P_C=60\% \)}
\end{figure}

We can see from the Fig.~\ref{fig:size1} that at the critical point the separation distance of Majorana pairs scales linearly with the system size; all system sizes collapse. This is  fractal behavior, the distribution of rare region size,  looks the same at all length scale. And the size distribution is very broad. large sizes have high probabilities.

\begin{figure}[htb]
	\includegraphics[width=\linewidth]{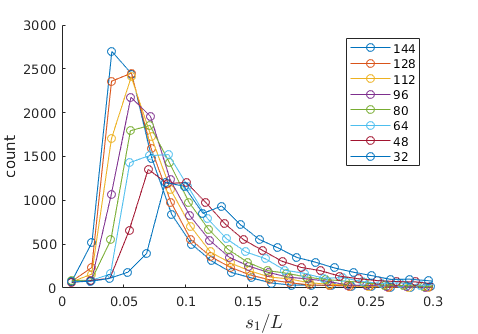}%
	\caption{\label{fig:size2}  the distribution of the size of Majorana pair seperation at the off the critical point \(P=40\% \)} 
\end{figure}

In contrast, in Fig.~\ref{fig:size2}  the size distribution  is plotted in the off-critical case. The distribution is very narrow, most of them is less than \(0.3\) of the system size. These don't collapse on the same distribution curve, as the system size increases; the distribution shifts to the left, which means that the relative size of rare regions are getting smaller and smaller. This does not mean that the rare regions will vanish in the thermodynamic limit \(L\rightarrow \infty \). The the size of the rare regions may still grow as \(s_1 \sim L^\theta \), but with \(\theta<1\). And the macroscopic number of zero modes will contribute to the non-universal power law behavior of the thermodynamics properties.

Now, the low energy mode in the previous chapter can be explained by the emergent Majorana modes.
The Eq.~\ref{eq:MZMsize} is much easier to calculate than the spin-spin correlation function,  and the rare region  information can be derived from the Majorana picture.

\section{Disorder with longer range interaction}\label{lrange}

\subsection{induced Majorana modes}

\begin{figure}[htb]
	\includegraphics[width=1\linewidth]{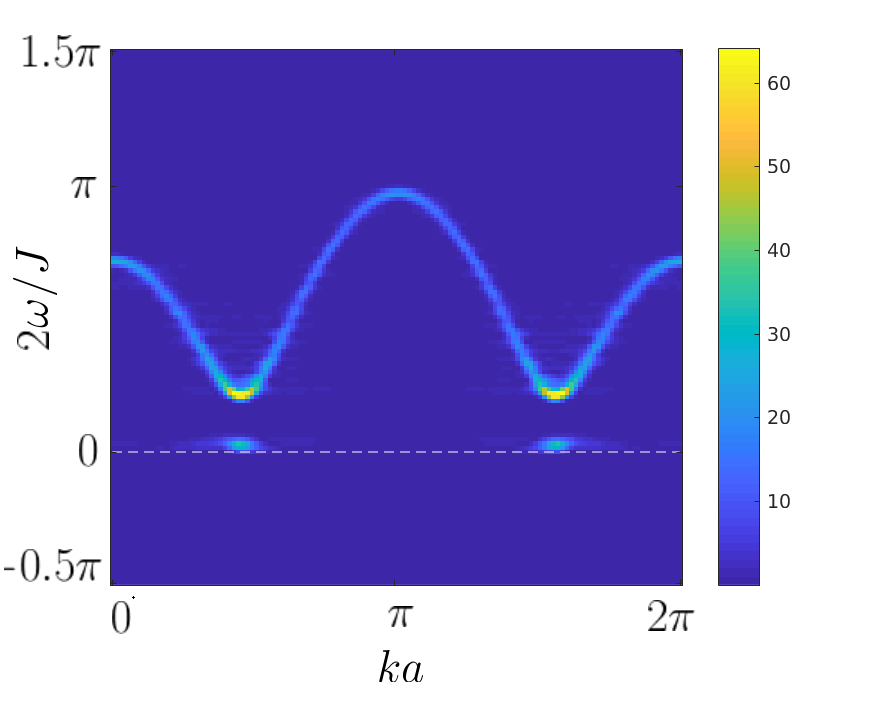}%
	\caption{\label{fig:3spinsB}   \(S(k,\omega)\)  for the binary distribution of the transverse field: \(P(h_L=1.6)=95\% \)  \(p(h_S=0.1)=5\%\);  here  \(\lambda_1=0.4\)  \(\lambda_2=-1.0 \)}
\end{figure}

\begin{figure}[htb]
	\includegraphics[width=1\linewidth]{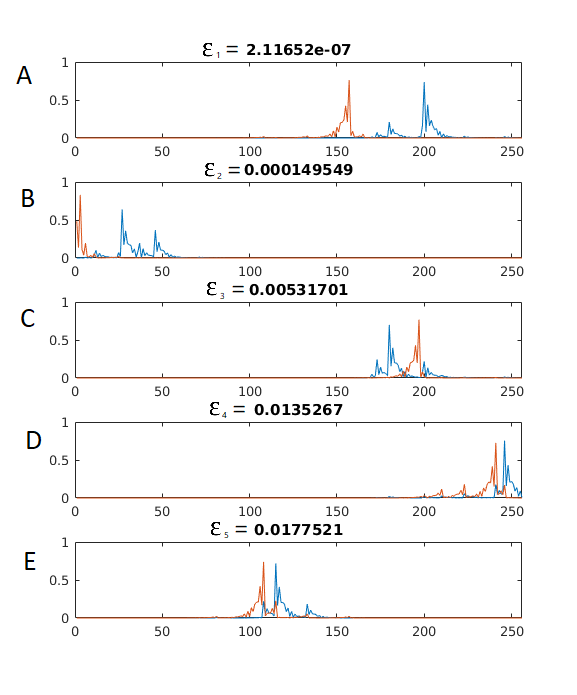}%
	\caption{\label{fig:AAAAA}    Five lowest eigenvectors, titles are the energy  \(\epsilon_n \); index \(n\) is in the ascending order of eigenenergy The horizontal axis is the lattice sites; the vertical axis is the value of \( \psi_{ni} \) and \(\phi_{ni}\). The orange and blue curves correspond to decoupled Majorana pairs: the  real part \( \tilde{a}_n=\sum_i\psi_{ni} a_i \) and  imaginary part \( \tilde{b}_n=\sum_i\phi_{ni} b_i \).  A,B,C,D,E  correspond to the rare regions in Fig.~\ref{fig:AAAAB}   }
\end{figure}

\begin{figure}[htb]
	\includegraphics[width=1\linewidth]{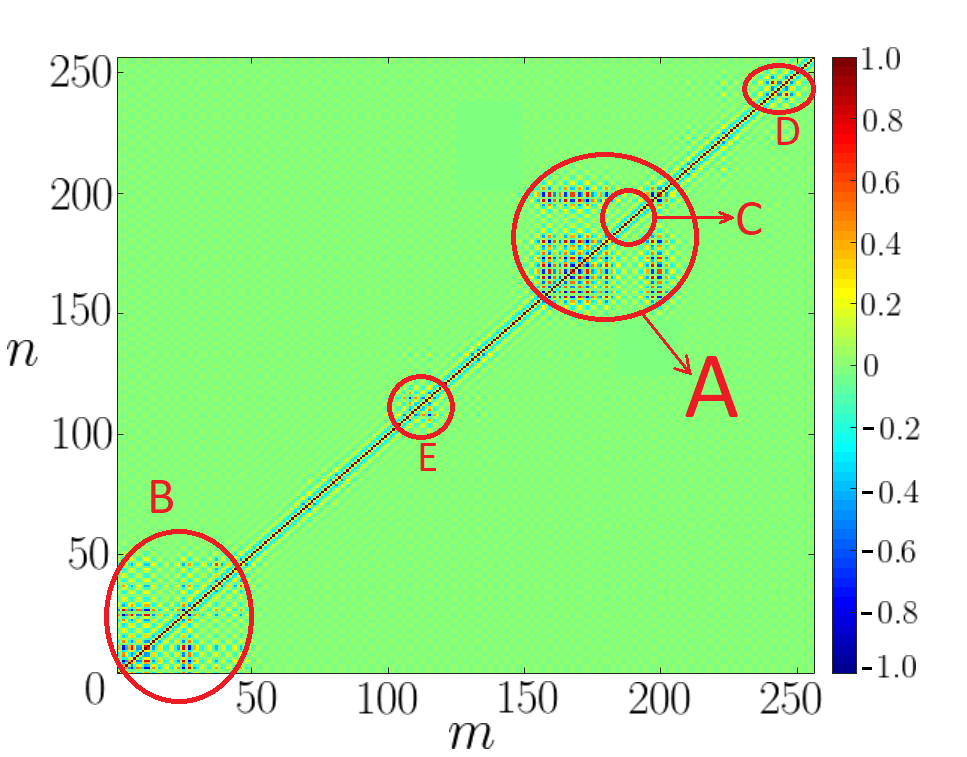}%
	\caption{\label{fig:AAAAB}  equal time spin-spin correlation \( <\sigma^x_n \sigma^x_m> \) ,  the horizontal axis is \(m\) the vertical axis is \(n\), the color is the strength of \(<\sigma^x_n \sigma^x_m> \)   }
\end{figure}

Let \(\lambda_1\) be non-zero. The results are given in Fig.~\ref{fig:3spinsB} and Fig.~\ref{fig:AAAAA}; the rare region diagram is shown in Fig.~\ref{fig:AAAAB}.
Due to the competition between \(\lambda_1\) and \(\lambda_2\), the Majorana zero mode oscillation pattern shifts to a new \(k\) value, between \(0\) and \( \pi \).  In Fig.~\ref{fig:3spinsB}, we can see the zero energy modes exist below the minimum of the dispersion curve. This is a very general phenomenon, the zero mode in the disordeed system is usually located where the pure system has  the smallest gap. The calculation is carried with the following Hamiltonian. 

\begin{eqnarray}
H=-\sum_{i=1}^L h_i \sigma^z_i -0.4 \sum_{i=1}^{L-1} \sigma^x_i \sigma^x_{i+1} +\sum_{i=1}^{L-2} \sigma^x_i \sigma^z_{i+1} \sigma^x_{i+2}    \nonumber \\
h_i=\begin{cases}
h_L=1.6  \qquad \text{probability}=95\%  \\
h_S=0.1 \qquad \text{probability}=5\% 
\end{cases}
\label{eq:3spinsB}
\end{eqnarray}
 Note that, the rare region $C$ is inside another rare region $A$ in Fig.~\ref{fig:AAAAA}.

\subsection{Spin glass phase}
In the three spin problem consider   setting
the transverse field and the next nearest coupling to be  \[ h=1  \qquad \lambda_2
=-0.3 \]
The  nearest couplings \(\lambda_{1i}\) are chosen to be random variables, satisfying the uniform distribution.
$[\lambda_1-\delta J, \lambda_1+ \delta J]$.
With \(\lambda_2\) providing frustration and \(\lambda_{1i}\) providing disorder, we expect to see a spin-glass phase transition as a function of \(\delta J\)

\begin{figure}[htb]
	\centering
	\includegraphics[width=0.5\textwidth]{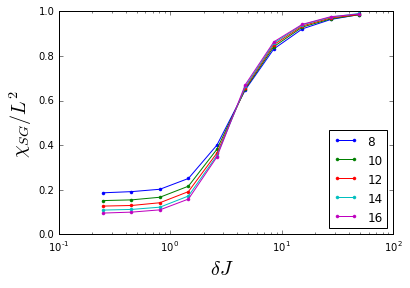}
	\caption{ \label{fig:L2} spin glass order for different system sizes. There is a phase transition near the disorder strength \(\delta J_c \approx 4\)}
\end{figure}

The spin glass [SG] order is defined by
\[
\chi_{SG}=\big[\sum_{i,j=1}^L \langle \sigma^x_i \sigma^x_j \rangle ^2\big]
\]
there are \(L^2\) terms in the summation, the square parenthesis corresponds to disorder average.
\begin{itemize}
	\item When  all sites are correlated, deep in the SG phase,
	\[\chi_{SG} \sim L^2 \]
	\item In the other extreme  case, non-SG phase, \(i\) and \(j\) are correlated only within some distance \(\xi\) \[\chi_{SG}\sim \xi L \]
\end{itemize}
In the Fig.~\ref{fig:L2}, we plot  \(\chi_{SG}/L^2\)

\begin{figure}[htb]
	\centering
	\includegraphics[width=0.5\textwidth]{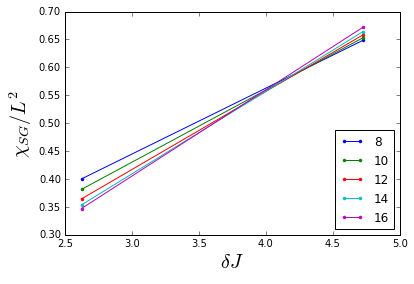}
	\label{fig:zoom}
	\caption{Zoom of the plot of the spin glass order Fig.~\ref{fig:L2}, the critical point is near \(\delta J_c\approx 4\)}
\end{figure}
    
\section{conclusion}

In this paper, we have explored quenched binary disorder in TFIM, and  a model   recently extended to include to contain  a three spin term. In the structure factor we find  strong zero energy modes whenever binary disorder straddles two distinct phases defined by winding numbers, analog of Griffiths-like rare region.  A previous attempt in TFIM to explain~\cite{PhysRevB.74.172414} this phenomenon was not satisfactory.   We show here from far more extensive calculations that it can be explained by MZM modes  induced by rare regions. The results also hold for the three-spin extended model.

The distribution of separation distance of the lowest energy Majorana mode pairs was defined. This quantity is very easy to calculate. We have shown that it has a fractal behavior at the critical point. The most probable size of Majorana modes is about 1/5th of the system size.

With the three spin interaction, the phase diagram becomes quite complex,~\cite{PhysRevB.85.035110} as was discussed previously.  lt is interesting that even in this case  the rare region induced MZM picture still holds, implying that our interpretation in terms of Griffiths-like rare region must have some validity -- note that topological orders are not defined by local order parameters. In the three spin model  a spin-glass phase transition is observed as a result of both frustration and disorder. 

Whether or not our work could be experimentally accessible remains problem for the future.

\begin{acknowledgments}
The authors would like to thank the condensed matter theory group for their patience through the various versions of this work. This work was supported in part by funds from the David S Saxon Presidential term Chair.
\end{acknowledgments}

\appendix
\section{diagonalizing a Hamiltonian with particle-hole symmetry}
\label{Appendix:A}
After the Jordan-Wigner transformation, we get a single particle Hamiltonian Eq.~ (\ref{eq:fermionHamiltonian}), which we can also rewrite it in a more compact Nambu basis \( \Psi^\dagger=(c^\dagger_1, \cdots, c^\dagger_L , c_1, \cdots, c_L) \)
\begin{eqnarray}
 H=\Psi^\dagger \begin{pmatrix} 
 A & B \\
 -B & -A 
 \end{pmatrix} \Psi
\end{eqnarray}
where \(A=\frac{1}{2}(M+M^T)\) and \(B=\frac{1}{2} (M-M^T) \)

\(M\) is an \(L\times L\) dimensional matrix, which contains all the information of the transverse fields and the couplings:
\begin{eqnarray}
M=\begin{pmatrix}
h_1 & -\lambda_{11} &  -\lambda_{21} &  & & &\\ 
& h_2 &   -\lambda_{12}&  -\lambda_{22} & & &\\ 
&  & h_3 &   -\lambda_{13}&  \ddots & &\\ 
&  &  & \ddots & \ddots& -\lambda_{2,L-2}\\ 
&  &  &  & \ddots&  -\lambda_{1,L-1}\\ 
&  &  &  & &  h_L\\ 
\end{pmatrix}
\end{eqnarray}

We can diagonalize the Hamiltonian kernel \(\begin{pmatrix} 
A & B \\
-B & -A 
\end{pmatrix} \) and get \(2L\) eigenvalues and eigenvectors. But this method doesn't take advantage of the particle-hole symmetry of the Hamiltonian kernel. That is, if \( \begin{pmatrix}
x \\ y
\end{pmatrix} \) is an eigenvector with eigenvalue \(\epsilon\), then  \( \begin{pmatrix}
y \\ x
\end{pmatrix} \) is also an eigenvector with eigenvalue \(-\epsilon\). 

For \(\epsilon\) close to zero, the \(\pm\epsilon\) pairs will have great relative error. If the system has multiple zero modes, the mixing error is even more complicated. Unfortunately, these zero Majorana modes are just what we are interested in. We need an new eigenvalue solver, taking advantage of the particle-hole symmetry.

The solution is to use the singular value decomposition of \(M\) (SVD).
\begin{eqnarray}
  M=\phi \Lambda \psi^T
\end{eqnarray}
the columns of \(\phi\) and \(\psi\) gives the coefficients in the Majorana representation, Eq,~(\ref{eq:majoranaHamiltonian}),

 \( \tilde{b}_n=\sum_i\phi_{in} b_i \) \   \( \tilde{a}_n=\sum_i\psi_{in} a_i \)   \  \( \epsilon_n=\Lambda_{nn} \)
 
\section{A numerical method to calculate Pfaffian}

We are using a very simple and effective method of calculating Pfaffian for any \(2N\times2N\) skew-symmetric matrix given in  Ref.~\onlinecite{PhysRevB.74.172414}.
Let \(X\) be a \(2N\times 2N\) skew-symmetric matrix, with the decomposition:

\begin{eqnarray}
	X=\begin{pmatrix}
	A & B\\ 
	-B^T & C
	\end{pmatrix}.
\end{eqnarray}
Then ($I_{n}$ is an identity matrix)
\begin{eqnarray}
 \begin{pmatrix} I_{2 } & 0\\ B^TA^{-1} & I_{2N-2} \end{pmatrix} X \begin{pmatrix} I_{2 } & -A^{-1}B\\ 0 & I_{2N-2} \end{pmatrix} =\begin{pmatrix} A & 0\\ 0 & C+B^TA^{-1}B \end{pmatrix}\qquad 
\end{eqnarray}
and
\begin{eqnarray}
	\det(X)=\det(A)\det(C+B^TA^{-1}B)
	\label{eq:pfaffian}
\end{eqnarray}
Equation (\ref{eq:pfaffian}) gives us a iteration method.
Each iteration, we find an \(A=\begin{pmatrix}
0 & a_{12} \\
-a_{12} & 0
\end{pmatrix}\) from the \(X\), such that \(|a_{12}|\) is the largest(for stability purposes).
Recalling that  \(\text{pf} \sim \sqrt{\det}\) up to an undetermined sign. However,  the sign of \(\text{pf}(A)=a_{12} \) is clear, so we have:
\begin{eqnarray}
\text{pf}(X)=a_{12}  \  \text{pf}(C+B^TA^{-1}B).
\end{eqnarray}
For the next iteration step, set:\(X' \rightarrow C+B^TA^{-1}B\), and repeat. We expect to see the final result to look like:
$$\text{pf}=a_{12} a'_{12} a''_{12} a'''_{12} a''''_{12} \cdots$$
Note that the matrix \(A\) doesn't have to be in the position shown in Equation (\ref{eq:pfaffian}); we can always trivially exchange the columns \(1\leftrightarrow i\) and rows  \(2\leftrightarrow j\) , making   \(A=\begin{pmatrix}
0 & a_{ij} \\
-a_{ij} & 0
\end{pmatrix}\) to be \(A=\begin{pmatrix}
0 & a_{12} \\
-a_{12} & 0
\end{pmatrix}\).

%
%

\end{document}